\documentclass[nofootinbib,prd,twocolumn,showpacs,showkeys,preprintnumbers]{revtex4-1}
\usepackage{hyperref,amssymb,amsmath,mathrsfs,bm,graphicx}
\usepackage[dvipsnames]{xcolor}
\begin{document}

\title {Complexity factor for black holes in the framework of the Newman-Penrose formalism}

\author{P. Bargue\~no}
\email{pedro.bargueno@ua.es}
\affiliation{Departamento de F\'{i}sica Aplicada, Universidad de Alicante, Campus de San Vicente del Raspeig, E-03690 Alicante, Spain}

\author{E. Fuenmayor}
\email{ernesto.fuenmayor@ciens.ucv.ve}
\affiliation{Centro de F\'isica Te\'orica y Computacional,\\ Escuela de F\'isica, Facultad de Ciencias, Universidad Central de Venezuela, Caracas 1050, Venezuela}

\author{E. Contreras }
\email{econtreras@usfq.edu.ec}
\affiliation{Departamento de F\'isica, Colegio de Ciencias e Ingenier\'ia, Universidad San Francisco de Quito,  Quito 170901, Ecuador.\\}

\begin{abstract}
In this work, we introduce the {\it complexity factor} in the context of self--gravitating fluid distributions for the case of black holes by employing the Newman-Penrose formalism. In particular, by working with spherically symmetric and static AdS black holes, we show that the complexity factor can be interpreted in a natural way at the event horizon. Specifically, a thermodynamic interpretation for the aforementioned complexity factor in terms of a pressure partially supporting a Van der Waals-like equation of state is given.
\end{abstract}


\maketitle

\section{Introduction}

Einstein field equations (EFE) constitute the basis of the General Relativity (GR) theory that establishes a relationship between the space--time geometry and the energy content in space. The geometry of space-time, which describes the gravitational field, is represented by the Einstein tensor, while the total energy content of space is described by the energy--momentum tensor. Of all the theories that gives us the understanding of how the gravitational field affects its vicinity, is undoubtedly the one that enjoys the greatest scientific acceptance. 

For instance, if we are seeking for non--vacuum solutions in the static, spherically symmetric and anisotropic case, the set of three independent field equations should be used to find five (metric and/or matter) functions, so in order to solve the system of equations, in this case we must provide two extra conditions \cite{LH-All-Static, EF-All-Static}. Many studies have been carried out in order to establish the appropriate conditions that, in addition to allowing the system to be closed, lead us to obtain real physical models that may be used to describe compact objects. Most of the conditions are given by means of state equations that intent to describe the local physics properties of the relativistic fluids what stars are made of, but there may be other possibilities, like plausible heuristic conditions over the metric variables which perfectly work to close and solve the system of field equations (see  \cite{EF-CP, LH-Poly-CP,Karmarkar,Cond, Cond1, Cond2, Cond3, Cond4, Cond5, Cond6, Cond7, Cond8, Cond9} and references therein). Following this scheme, we can use the relevant concept of complexity (in the realm of general relativity) for self-gravitating relativistic fluids, as a condition that provides extra information which can categorize the system.\\

The concept of complexity in physics is anything but intuitive, so defining the complexity of various (some how, different) systems is not a trivial task. This concept is related to many intriguing aspects deepen in the structure of nature, therefore, it has been carefully studied by many researchers and applied to different scenarios in science \cite{complex1,comp1, comp2, comp3, comp4, comp5, comp6, comp7, comp8, comp9, comp10, comp11, comp12, comp13, comp14, comp15, comp16, comp17, comp18}. However, to this day, we do not enjoy of consensus on how to define complexity for the many edges that nature has. In physics, various efforts have been devoted to provide a clear definition of complexity which was originally related to aspects like information and entropy \cite{comp11, comp12, comp13} based on the idea of measuring the basic internal structure of a system. In this regard, when dealing with an intuitively ``complex'' system, we should have the ability to define an observable capable to measure and quantify its complexity, and thus be able to discriminate different systems due to their degree of complexity. Establishing a hierarchy that catalogs systems according to their complexity and that would make associated physical characteristics flourish would be a relevant task \cite{Bondi}.

In the case of general relativistic compact objects, recently, a more intuitive definition of complexity was established based on the structure of the relativistic fluid itself: inhomogeneity of the energy density (density constrast) and pressure anisotropy. In principle, these characteristics are not directly related to information. Particularly, the new approach proposed by Herrera \cite{complex1} was based on a definition for the complexity factor (denoted by $Y_{TF}$) that is manifestly related to the internal structure of the fluid distribution. The structure scalar $Y_{TF}$, that exhibits the complexity of a gravitational system, arises from the orthogonal splitting of the Riemann tensor \cite{Bel,LH-C3} and was initially defined for static spherically symmetric relativistic anisotropic fluids \cite{complex1}. The basic assumption of this proposal is sustained by assuming that one of the less complex (simpler) systems corresponds to an homogeneous fluid with isotropic pressure, so a zero complexity value is assigned for such a distribution. In this regard the vanishing complexity condition can be viewed like a non–local equation of state used to obtain other non-trivial compact configurations with zero complexity \cite{complex1}. It is worth emphasizing that, although the definition of complexity was introduced to deal with interior solutions, this concept has been extended to vacuum solutions that are represented by the Bondi metric, where a kind of complexity hierarchy is established, ranging from the Minkowski space--time (vanishing complexity) to gravitationally radiating systems (more complex) \cite{Bondi}. Therefore, in this work we extrapolate such a definition for black hole (BH) geometries with the aim of shedding some light on a possible relation between the complexity factor, $Y_{TF}$, and the BH thermodynamics via the Newman-Penrose (NP) formalism. At this point we would like to emphasize that, the definition of complexity used in our work is associated to the very concept of ``structure'' within the fluid distribution and is not related with information, accessible states or possible paths of states or quantum gates (at least not directly).


This work is organized as follows. In the next section we set up all the equations and conventions to deal with the system of Einstein field equations. Its presented in Section 3 a brief revision of the definition for the complexity factor for a static--spherically symmetric--anisotropic relativistic fluid distribution. Next, the conditions for extrapolating the definition of the complexity factor for the case of black hole geometries is displayed in section 4. Section 5, is dedicated to the interpretation of the role played by the complexity factor in black hole thermodynamics. In the last section, some final remarks are presented.

\section{General Relativity relevant equations}\label{Einstein}

Here we will focus on a static fluid with spherical symmetry. This fluid is taken to be anisotropic and bounded by a surface denoted by $\Sigma$.
The corresponding line element, written using Schwarzschid-like {\it ansantz}, is
\begin{eqnarray} \label{metrica}
 ds^2 = e^{\nu(r)} dt^2 - e^{\lambda(r)} dr^2 - r^2 \left( d\theta^{2} + \sin^{2}\theta d\phi^{2}\right),
\end{eqnarray}
where $\nu(r)$ and $\lambda(r)$ are functions depending only on the radial coordinate, $r$.
The aforementioned metric satisfies Einstein field equations (EFE)
given by \footnote{We are assuming geometric units $c=G=1$}
\begin{eqnarray} \label{EFE}
 G^{\alpha}_{\beta} = 8 \pi T^{\alpha}_{\beta}.
\end{eqnarray}
The matter sector is described by the following
energy-momentum tensor:
\begin{eqnarray}\label{energia-momentum}
T_{\alpha\beta}=(\rho+P_{\perp})u_{\alpha}u_{\beta}-P_{\perp}g_{\alpha\beta}+(P_{r}-P_{\perp})s_{\alpha}s_{\beta},
\end{eqnarray}
where, 
\begin{eqnarray}\label{4-velocity}
u^{\alpha}=(e^{-\nu(r)/2},0,0,0),
\end{eqnarray}
is the four velocity of the fluid and $s^{\alpha}$ is defined as
\begin{eqnarray}
s^{\alpha}=(0,e^{-\lambda(r)/2},0,0),
\end{eqnarray}
with the properties $s^{\alpha}u_{\alpha}=0$, $s^{\alpha}s_{\alpha}=-1$. Using the four-velocity (\ref{4-velocity}), the four acceleration can be written as $a^{\alpha}=\nabla_\beta (u^{\beta})u^{\alpha}$ which, in this case, is given by a single component
\begin{eqnarray}\label{4-acele}
a_{1}=-\frac{\nu '}{2}.
\end{eqnarray}
where the prime indicates partial derivation with respect to the $r$ coordinate.
It will be convenient to rewrite the energy-momentum tensor (\ref{energia-momentum}) like \cite{complex1},
\begin{eqnarray}\label{energia-momentum2}
T^{\alpha}_{\beta}=\rho u^{\alpha}u_{\beta}-P h^{\alpha}_{\beta}+ \Pi^{\alpha}_{\beta} ,
\end{eqnarray}
with
\begin{eqnarray} \label{OS2}
&&\Pi^{\alpha}_{\beta}=\Pi  \left(s^{\alpha}s_{\beta} + \frac{1}{3} h^{\alpha}_{\beta} \right) ; \quad P=\frac{P_{r}+ 2P_{\perp}}{3}\nonumber\\
 &&\quad\quad\;\;\; \Pi = P_{r} - P_{\perp}; \quad\quad\;\; h^{\alpha}_{\beta} = \delta^{\alpha}_{\beta}-u^{\alpha}u_{\beta}.
\end{eqnarray}

For the static case, EFEs (\ref{EFE}) are given by
\begin{eqnarray}
\rho&=&-\frac{1}{8\pi}\bigg[-\frac{1}{r^{2}}+e^{-\lambda}\left(\frac{1}{r^{2}}-\frac{\lambda'}{r}\right) \bigg],\label{ee1}\\
P_{r}&=&-\frac{1}{8\pi}\bigg[\frac{1}{r^{2}}-e^{-\lambda}\left(
\frac{1}{r^{2}}+\frac{\nu'}{r}\right)\bigg],\label{ee2}
\end{eqnarray}
\begin{equation}
P_{\perp}=\frac{1}{8\pi}\bigg[ \frac{e^{-\lambda}}{4}
\left(2\nu'' +\nu'^{2}-\lambda'\nu'+2\frac{\nu'-\lambda'}{r}
\right)\bigg]\label{ee3},
\end{equation}
where primes denote derivative with respect to $r$. 

From the radial component of the conservation law (or using Einstein's equations)
\begin{equation}
    \nabla_\alpha T^{\alpha \beta} = 0,
\end{equation}
one can obtain the generalized Tolman-Opphenheimer--Volkoff (hydrostatic equilibrium) equation for anisotropic matter which reads
\begin{equation}
P'_r=-\frac{\nu'}{2}\left(\rho + P_r\right)+\frac{2\left(P_\bot-P_r\right)}{r}.\label{Prp}
\end{equation}
Alternatively, using 
\begin{equation}
\nu' = 2 \frac{m + 4 \pi P_r r^3}{r \left(r - 2m\right)},
\label{nuprii}
\end{equation}
which can be derived from EFEs, we arrive to \eqref{Prp} as
\begin{equation}
P'_r=-\frac{(m + 4 \pi P_r r^3)}{r \left(r - 2m\right)}\left(\rho + P_r\right)+ \frac{2\left(P_\bot-P_r\right)}{r},\label{ntov}
\end{equation}
where the mass function, $m$, is defined by
\begin{equation}
R^3_{232}=1-e^{-\lambda}=\frac{2m}{r},
\label{m}
\end{equation}
or, equivalently
\begin{equation}
m= 4\pi \int_{0}^{r} \tilde{r}^{2} \rho d\tilde{r}.
\label{m2}
\end{equation}

Although Tolman's mass \cite{Tolman} was introduced as a measure of the total energy of the system, where the integration is down to the radius surface $r_{\Sigma}$, we shall define the mass within a sphere of radius $r$ (completely inside $\Sigma$) for a spherically symmetric static distribution of matter as
\begin{equation}
m_{T}= 4\pi \int_{0}^{r} \tilde{r}^{2} e^{(\nu + \lambda)/2} (T^{0}_{0} - T^{1}_{1} - 2 T^{2}_{2}) d\tilde{r}.
\label{mTW}
\end{equation}


We end this section by writing another expression for Tolman's mass $m_T$ that will be useful in establishing physical relationships and interpretations (see references \cite{LH-Report,LH-Aniso} for details and for a complete discussion),
\begin{eqnarray}
&&m_{T}= (m_{T})_{\Sigma} \left(\frac{r}{r_{\Sigma}}\right)^3
- r^3 \int_{r}^{r_{\Sigma}} e^{(\nu + \lambda)/2} 
\left[ \frac{8\pi}{\tilde{r}} (P_{\perp}-P_{r}) \right] d\tilde{r}
\nonumber\\
&&- r^3 \int_{r}^{r_{\Sigma}} e^{(\nu + \lambda)/2} 
\left[\frac{1}{\tilde{r}^4} \int_{0}^{\tilde{r}} 4\pi \tilde{r}^{3} \rho ' d\tilde{r} \right] d\tilde{r}.
\label{mTW5}
\end{eqnarray}
As we see in (\ref{mTW5}), the second term in the right side describes the contribution of density inhomogeneity and local anisotropy of pressure to Tolman's ``active gravitational'' mass. We will see next that these properties enter into the definition of complexity for relativistic fluids.\\


\section{Complexity Factor for self--gravitating spheres}\label{complexityF}

Recently, Herrera has introduced an interesting
definition of complexity of self--gravitating fluids \cite{complex1}.
In this section, we summarize the essential aspects of this definition for the complexity factor which is based mainly on
the following idea: a homogeneous energy density distribution supported by an isotropic pressure possesses a minimal complexity.

The starting point of the discussion will be the well known orthogonal splitting of the Riemann tensor expressed in terms of physical variables. We begin by stating that the Rieman curvature can be expressed through the Weyl tensor $C^{\rho}_{\alpha \beta\mu}$, the Ricci tensor $R_{\alpha\beta}$ and the curvature scalar $R$, by
\begin{eqnarray} \label{Riemann1}
 R^{\rho}_{\alpha\beta\mu}&=& C^{\rho}_{\alpha \beta\mu} + \frac12 R^{\rho}_{\beta} g_{\alpha\mu} - \frac12 R_{\alpha\beta} \delta^{\rho}_{\mu}+\frac12 R_{\alpha\mu}\delta^{\rho}_{\beta}\nonumber\\&-&\frac12 R^{\rho}_{\mu} g_{\alpha\beta}-\frac16 R\left(\delta^{\rho}_{\beta}g_{\alpha\mu} - g_{\alpha\beta}\delta^{\rho}_{\mu}\right).
\end{eqnarray}
Due to the lack of rotation, the magnetic part of the Weyl tensor vanishes and, therefore, only the electric part will be present as
\begin{eqnarray} \label{Weylelectrico}
 E_{\alpha\beta} = C_{\alpha\gamma\beta\delta}u^{\gamma}u^{\delta}.
\end{eqnarray}
Note that $E_{\alpha\beta}$ may also be written as \cite{LH-C3}
\begin{eqnarray} \label{E}
E_{\alpha \beta} = E \left(s_{\alpha}s_{\beta} + \frac{1}{3} h_{\alpha\beta}\right),
\end{eqnarray}
with
\begin{eqnarray} \label{E2}
E=-\frac{e^{-\lambda}}{4}\left[\nu '' + \frac{\nu '^{2} -\lambda ' \nu '}{2}- \frac{\nu ' - \lambda '}{r} + \frac{2(1-e^{\lambda})}{r^2}\right],\nonumber\\
\end{eqnarray}
form where it can be seen that
$E_{\alpha \beta}$ satisfies the following properties
\begin{eqnarray} \label{E3}
E^{\alpha}_{\;\alpha} = 0 , \qquad  E_{\alpha\gamma} = E_{(\alpha\gamma)}, \qquad E_{\alpha\gamma}u^{\gamma} = 0.
\end{eqnarray}
Further, using Einsteins equations (\ref{EFE}), the decomposition of the Riemann tensor (\ref{Riemann1}) and (\ref{E}) in the definition of the mass function (\ref{m}), we may write
\begin{eqnarray} \label{m3}
m = \frac{4\pi}{3} r^3 (\rho + P_{\perp} - P_{r} ) + \frac{r^3 E}{3},
\end{eqnarray}
from which it is simple to obtain
\begin{eqnarray} \label{E4}
E = - \frac{4\pi}{3} \int_{0}^{r}\tilde{r}^3 \rho '  d\tilde{r} + 4\pi (P_{r}-P_{\perp}).
\end{eqnarray}

Now, it can be shown \cite{Bel, LH-C2} that the Riemann tensor may be expressed through the tensors 
\begin{eqnarray} \label{OS1}
Y_{\alpha \beta}&=& R_{\alpha\gamma\beta\delta} u^{\gamma} u^{\delta}\\
Z_{\alpha \beta}&=& ^\ast\; R_{\alpha\gamma\beta\delta} u^{\gamma} u^{\delta}= \frac12 \eta_{\alpha\gamma\epsilon\mu}R^{\epsilon\mu}_{\;\;\;\;\beta\delta} u^{\gamma} u^{\delta}\\
X_{\alpha \beta}&=& ^\ast\; R^{\ast}_{\alpha\gamma\beta\delta} u^{\gamma} u^{\delta}= \frac12 \eta_{\alpha\gamma} ^{\;\;\;\;\epsilon\mu}R_{\epsilon\mu\beta\delta}^{\ast} u^{\gamma} u^{\delta}
\end{eqnarray}
in what is called the orthogonal splitting of the Riemann tensor \cite{Bel}. Here $\ast$ denotes the Hodge operator, {\it i. e.},   $R^{\ast}_{\alpha\beta\gamma\delta}=\frac12 \eta_{\epsilon\mu\gamma\delta} R_{\alpha\beta}^{\quad\epsilon\mu}$ and $\eta_{\epsilon\mu\gamma\delta}$ corresponds to the Levi--Civita tensor. After introducing the dynamical content of EFEs in this orthogonal splitting, we arrive to
(see \cite{LH-C3} for details of the general non–static case),
\begin{equation} \label{Riemann2}
 R^{\alpha\gamma}_{\;\;\;\;\beta\delta}= C^{\alpha\gamma}_{\;\;\;\;\beta\delta} + 16\pi T^{[\alpha}_{\;\;[\beta} \delta^{\gamma]}_{\;\;\delta]} + 8 \pi T \left(\frac{1}{3} \delta^{\alpha}_{\;\;[\beta} \delta^{\gamma}_{\;\;\delta]} - \delta^{[\alpha}_{\;\;[\beta} \delta^{\gamma]}_{\;\;\delta]}  \right).
\end{equation}
Then, replacing  (\ref{energia-momentum2}) in (\ref{Riemann2}) we can split the Riemann tensor in four characteristic parts that allow to obtain explicit expressions for the tensors $Y_{\alpha \beta}$, $Z_{\alpha \beta}$ and $X_{\alpha \beta}$ in terms of the physical variables. So, after some manipulations (see \cite{LH-C3}) we obtain
\begin{eqnarray} 
&&Y_{\alpha \beta} = \frac{4\pi}{3} (\rho + 3 P) h_{\alpha \beta} + 4\pi \Pi_{\alpha\beta} + E_{\alpha\beta}\label{Y}\\
&&Z_{\alpha \beta} = 0\\
&&X_{\alpha \beta} = \frac{8\pi}{3} \rho h_{\alpha\beta} +  4\pi \Pi_{\alpha\beta} - E_{\alpha\beta}\label{X}.
\end{eqnarray}

From the above expressions we can define the structure scalars (the scalar associated to the tensor $Z_{\alpha\beta}$ vanishes in the static case) \cite{LH-C3}
\begin{eqnarray} 
X_{T} &=& 8\pi \rho\label{XT}\\ 
X_{TF} &=&  4\pi \Pi - E \label{XTF}\\
Y_{T} &=& 4\pi \left(\rho  + 3 P_{r} -  2 \Pi \right)\label{YT}\\
Y_{TF} &=& 4\pi \Pi + E \label{YTF}.
\end{eqnarray}
Now, it follows immediately that local anisotropy of pressure is determined by $X_{TF}$ and $Y_{TF}$ by
\begin{eqnarray} \label{XTF+YTF}
X_{TF} + Y_{TF} =  8 \pi \Pi.
\end{eqnarray}
Using (\ref{E4}) it is straightforward to write $X_{TF}$ as
\begin{eqnarray} \label{XTF2}
X_{TF} = \frac{4\pi}{r^3}\int^{r}_{0} \tilde{r}^3 \rho' d\tilde{r},
\end{eqnarray}
Therefore, in the static and no dissipative case, $X_{TF}$ controls the inhomogeneities of the energy density.
These inhomogeneities have been studied, for instance, in the formation of naked singularities \cite{Newman} and in relation with Penrose's proposal on the gravitational arrow of time based on the Weyl tensor \cite{PenroseL} (in this sense, $X_{TF}$ should be the essential ingredient in the definition of a gravitational arrow of time). 
See (\cite{LH-C3}) for a full discussion related to this topic.

Here our main interest is focused on the interpretation of the complexity factor for black hole geometries. Directly, introducing (\ref{E4}) in (\ref{YTF}) we get
\begin{eqnarray} \label{YTF2}
Y_{TF} = 8\pi \Pi - \frac{4\pi}{r^3}\int^{r}_{0} \tilde{r}^3 \rho' d\tilde{r},
\end{eqnarray}
that allows us to express $Y_{TF}$ in terms of the inhomogeneity of the energy density (density contrast) and the local anisotropy of the system. An equivalent relation can be stated by noting that Eq. (\ref{YTF2}) allows us to write Tolman's mass \cite{LH-Report, LH-Aniso} as
\begin{eqnarray} \label{m_T}
m_{T} = (m_{T})_{\Sigma}\left(\frac{r}{r_{\Sigma}}\right)^3 + r^3\int^{r_{\Sigma}}_{r} \frac{e^{( \nu + \lambda )/2}}{{\tilde{r}}} Y_{TF} d\tilde{r},
\end{eqnarray}
which explicity shows that the complexity factor includes inhomogeneities and anisotropies of the energy density and pressure, respectively, on the active gravitational mass \cite{complex1}
Finally, we note that EFEs imply that Eq. (\ref{YTF2}) can be written as
\begin{eqnarray}\label{ytfcal}
Y_{TF}=\frac{e^{-\lambda } \left(\nu ' \left(r \lambda '-r \nu '+2\right)-2 r \nu ''\right)}{4 r},
\end{eqnarray}
which is written as a function of the metric potentials only.


\section{Complexity Factor for black hole-like geometries}

In this section, our purpose is to extrapolate the definition of the complexity factor for the case of BH geometries. In particular, we are interested in shedding some light on a possible relation between $Y_{TF}$ and BH thermodynamics {\it via} the NP formalism. Our motivation is the following: although it is clear that we can use the laws of BH dynamics to give a thermodynamics interpretation of the NP scalars \cite{Hayward}, it is not straightforward to identify the pressure within this context. While some prescriptions for the identification of pressure can be found in the literature (mainly due to the cosmological constant in the extended phase space approach \cite{Kubiznak} or to the matter content in horizon thermodynamics \cite{Padma}), the NP formalism was implemented in order to study the thermodynamic behaviour of some standard BH spacetimes in order to shed light into a more geometrical description of the relevant thermodynamic variables.

Let us fix our attention in geometries satisfying the Schwarzschild ansatz, $\nu'+\lambda'=0$, by writing the metric as
\begin{equation}
\label{metric}
ds^2=fdt^2-f^{-1}dr^2-r^2 d\Omega^2,
\end{equation}
where f is a function of the $r$ coordinate.

Now, after choosing the following null tetrad
\begin{eqnarray}
l^{\mu}&=&\left(1,f,0,0\right), \nonumber \\
n^{\mu}&=&\left(\frac{1}{2f},-\frac{1}{2},0,0\right), \nonumber \\
m^{\mu}&=&\left(0,0,\frac{1}{\sqrt{2}r},\frac{\textrm{i}\csc \theta}{ \sqrt{2}r}\right),
\end{eqnarray}
where $l_{\mu}n^{\mu}$=1 and $m_{\mu}\bar m^{\mu}=-1$ and the bar denotes complex conjugation. The only non--vanishing Weyl and Ricci scalars for the Einstein-non-null-Maxwell system (including nonlinear generalizations) (which will be either Petrov type D or O due to spherical symmetry) are 
\begin{eqnarray}
\Psi_{2}&=&C_{pqrs}l^{p} m^{q} \bar m^{r} n^{s} \\
\Phi_{11}&=&-\frac{1}{2}R_{ab}l^{a}n^{b}+3 \Lambda \\
\Lambda&=&\frac{R}{24}.
\end{eqnarray}
\\
We remind the reader that $\Psi_{2}$ is a Coulomb-like term which determines tidal forces, $\Phi_{11}$ is roughly the energy density of the matter sector and the $\Lambda$ term is essentially the Ricci scalar. By using Eq. (\ref{metric}), the above can be written as
\begin{eqnarray}
\Psi_{2}&=&-\frac{1}{6 r^2}+\frac{f}{6 r^2}-\frac{f'}{6 r}+\frac{f''}{12} \label{NPa} \\
\Phi_{11}&=&\frac{1}{4 r^2}-\frac{f}{4 r^2}+\frac{f''}{8} \label{NPb} \\
\Lambda&=&\frac{1}{12 r^2}-\frac{f}{12 r^2}-\frac{f'}{6 r}-\frac{f''}{24},\label{NPc}
\end{eqnarray}
where $\Lambda=\frac{R}{24}$. Interestingly, by using (\ref{m}), 
(\ref{NPa}), (\ref{NPb}) and (\ref{NPc}), the Misner--Sharp mass can be written as
\begin{equation}
\label{Misner}
m=(\Phi_{11}-\Psi_{2}+\Lambda)r^3.
\end{equation}
Using this result and combining (\ref{m3}) and (\ref{YTF}), we get that the complexity factor reads
\begin{equation}
    Y_{TF}=3\left(\Phi_{11}-\Psi_{2}+\Lambda\right)-4\pi\left(\rho-2\Pi\right).
\end{equation}
Now, in order to express the complexity factor in terms of the NP symbols only, we proceed as follows.
First, note that for a metric satisfying the Schwarzschild condition, Eq. (\ref{ytfcal}) takes the form
\begin{eqnarray}
\label{YnonNP}
Y_{TF}=\frac{f'-r f''}{2 r}.
\end{eqnarray}
Then, using EFE (\ref{EFE}), we get
\begin{equation}
    4\pi \left( \rho-2\Pi\right)=\frac{3}{2r^2}-\frac{3f}{2r^2}-\frac{f'}{2r}+\frac{f''}{2}.
\end{equation}
Finally, using the following relations which can be easily derived from Eqs. (\ref{NPa})-(\ref{NPc})
\begin{eqnarray}
\label{f1}
f&=&1-2r^2\left(\Lambda+\Phi_{11}-\Psi_{2}\right) \\
\label{fp}
f'&=&-2r\left(2\Lambda+\Psi_{2}\right)\\
\label{fpp}
f''&=&4\left(-\Lambda+\Phi_{11}+\Psi_{2}\right),
\end{eqnarray}
we obtain
\begin{equation}
        4\pi \left( \rho-2\Pi\right)=3\Lambda+5\Phi_{11},
\end{equation}
and therefore
\begin{equation}
\label{main1}
    Y_{TF}=-3\Psi_{2}-2\Phi_{11}, 
\end{equation}
which coincides with Eq. (\ref{YnonNP}) when the corresponding NP symbols are explicitly written in terms of the metric potential.

The expression given by Eq. (\ref{main1}) is one of the main results of the present work so a couple of comments regarding it are compulsory. First, note that  for an uncharged ($\Phi_{11}=0$) AdS BH, the complexity factor is non vanishing but, in the pure AdS case, it vanishes due to conformal flatness. Therefore, the complexity of these two states is not the same: pure AdS and charged AdS black holes have different complexity. Second, note that as $Y_{TF}$ is written in terms of the only non-vanishing NP scalars of non-rotating horizons, we speculate on the possibility that a general definition for non spherically symmetric (but Petrov-type D solutions) could be done by introducing the appropriate NP symbols which accounts for frame-dragging. Along the next section, we will elaborate on the possible meaning of $Y_{TF}$ in the context of BH thermodynamics.

\section{Interpretation of the complexity factor for static black hole geometries}

With the aim to elucidate the link between the complexity factor and BH thermodynamics, let us briefly review a direct approach towards it through the BH--Van der Waals model (see, for example, \cite{Vargas1,Vargas2} and references therein). In this regard, let us consider Einstein's equations for a Schwarzschild-like ansatz:
\begin{eqnarray}
\label{eqe1}
\frac{1-f}{r^2}+\lambda-\frac{f'}{r}&=&\rho \\
\label{eqe2}
-\frac{1}{2}f''+\lambda-\frac{f'}{r}&=&-p,
\end{eqnarray}
where $\lambda$ stands for a possible cosmological constant. 

Let us consider what happens on the event horizon (which for simplicity we will consider a single one). For this purpose, we remark that the conserved charge due to the Killing vector field $\partial_{t}$ is the Komar energy which, for this kind of geometry, is written as $E_{K}=\frac{1}{2}r^2 f'$. 
It is interesting also to note that, following Padmanabhan \cite{Padma2}, that $E_{K}$ can be written as an energy equipartition of the horizon degrees of freedom as $E_{K}\hat =\frac{1}{2}N T$, where $N=A=4\pi r^2/l_{p}^2$ and $T$ stands for the corresponding Hawking temperature (the $\hat=$ symbol denotes that the equality is valid only at the event horizon). As usual in literature, an euclidean (thermodynamic) volume, $V\hat=4 \pi r^3/3$, will be considered. In addition, if the cosmological constant is taken as a pressure, $P_{\lambda}=\frac{\lambda}{8\pi}$, then $V$ and $P_{\lambda}$ are conjugates when the horizon degrees of freedom are re-scaled as $\bar N\hat=N/6$. With these definitions, Eq. (\ref{eqe1}) reads $(8 \pi =1)$:
\begin{eqnarray}\label{eso}
    \frac{\bar N T}{V}&\hat=& \lambda-k_{g}-\rho, 
\end{eqnarray}
where $k_{g}$ corresponds to the  Gaussian curvature at the horizon. 
Note that, after considering a charged AdS BH,
Eq. (\ref{eso}) leads
 \cite{Vargas1}:
\begin{equation}
\label{VdW}
    P_{\lambda}\hat=\frac{\bar N T }{V}-\frac{1}{2\pi}\frac{\bar {N}^2}{V^2}+\frac{2 Q}{\pi}\frac{\bar{N}^4}{V^4},
\end{equation}
which corresponds to equation of state of a Van der Waals gas
where the {\it interaction} and {\it second virial} term become manifest, being $Q$ the electric charge.

As commented before, a nice geometrization of this equation of state \cite{Villalba} can be set using the NP formalism as follows. In Ref. \cite{PRbook}, Penrose and Rindler defined the quantity
\begin{equation}
\label{K}
K=-\sigma \lambda -\Psi_{2}-\tilde \rho \mu +\Phi_{11} + \Lambda ,
\end{equation}
as the complex curvature for arbitrary two--surface in spacetime which are spacelike. In (\ref{K}),
$\sigma$, $\lambda$, $\tilde \rho$ and $\mu$ are spin coefficients related to the expansion and shear of the null vectors
$l$ and $n$. Additionally, in \cite{PRbook} it is shown that
the Gaussian curvature of such a two--surface can be written as 
\begin{equation}
\label{gauss}
K+\bar K=k_{g}
\end{equation}
For example, for a charged AdS BH (which has a shear-- and expansion--free horizon), Eq. (\ref{gauss}) reads
\begin{equation}
\label{Delta}
\frac{\Delta}{2r^{4}}\hat=\frac{k_{g}}{2}+ \Psi_{2}-\Phi_{11}-\Lambda =0,
\end{equation}
where $\Delta = r^2-2 M r + Q^2-\lambda r^4/3$.
\\
At this point, the complexity factor can be introduced in two equivalent ways. First of all, note that Eq. (\ref{eqe2}) can be trivially written in terms of $Y_{TF}$. The interesting point to observe is that Eq. (\ref{eqe2}) also contains the $\frac{f'}{r}$-term, which, when evaluated at an event horizon, results in the aforementioned {\it kinetic} term $\frac{\bar N T}{V}$ together with an explicit pressure due to the matter sector, $p$. Therefore, $Y_{TF}$ can be included in some kind of equation of state associated with the event horizon.

The second, more geometrical way of interpreting $Y_{TF}$, can be established with the help of Eqs. (\ref{f1})-( \ref{fpp}) and (\ref{main1}). Specifically, those expressions allow us to write Eqs. (\ref{gauss}) or (\ref{Delta}) as
\begin{equation}
\label{main}
    \frac{Y_{TF}}{4\pi}\hat=\frac{\bar N T}{V}-\frac{1}{\pi}\frac{\bar N^2}{V^2}-\frac{\Phi_{11}}{4\pi},
\end{equation}
where $\frac{\Phi_{11}}{4\pi}$ is the pressure due to the matter sector (in fact, $\Pi=4\Phi_{11}$).
\\
\\
Comparison of Eqs. (\ref{VdW}) and (\ref{main}) reveals, by inspection, some interesting connections. For instance, Eq. (\ref{main}) shows that the complexity factor is what sustains the {\it formal gas}. In this sense, the event horizon corresponds to the null surface where the $Y_{TF}$, the kinetic term, the horizon curvature term (the Van der Waals-like like term) and the radiation pressure, are in equilibrium. Therefore, a vanishing {\it complexity} means, in this context, that the gas is only supported by kinetic, horizon and Maxwellian terms.

\section{Conclusions}

It is well known the fact that it can be associated thermodynamic properties with black hole horizons and different attempts have been made to relate aspects linked to the internal gravitational thermodynamic structure of dense objects with the principles of quantum theory. This could possibly offer an open window to the understanding of quantum geometry. These attempts have brought together some features like path integral formulations for the study of horizon temperature and black hole evaporation, or also current and very interesting aspects related to bifurcation horizons, concepts of Noether charge and the relation with horizon entropy \cite{Kubiznak, Padma2}. In the present work, we followed an alternative route extrapolating the new definition of the complexity factor for self-gravitating spheres (which is one of the scalars that arise from the orthogonal splitting of the Riemann tensor) to the black hole geometry domains. In particular, we demonstrated that the complexity factor, $Y_{TF}$, can be expressed only in terms the Newman-Penrose symbols which have played a fundamental role in the connection between geometry and thermodynamics of black holes.  In this sense, we have obtained a formal connection between the complexity factor and black hole thermodynamics, which could be interesting to be extended to rotating solutions. Also, including the interpretation for other structure scalars (like $X_{TF}$ for example) within the present formalism may be interesting to consider in future developments.

\acknowledgments{P. B. acknowledges Anaís, Lucía, Inés and Ana for continuous support. P. B. is funded by the Beatriz Galindo contract BEAGAL 18/00207 (Spain).}

\end{document}